\documentclass[namedreferences]{solarphysics}
\usepackage[breaklinks=true,unicode=true,pdfborder={0 0 0}]{hyperref}
\usepackage{breakurl}
\hypersetup{urlcolor=blue, linkcolor=black}
\usepackage[optionalrh,solaromanenum]{spr-sola-addons} 
\usepackage{graphicx}                    
\usepackage{color}                       
\usepackage{url}                         

\begin{document}
\bibliographystyle{spr-mp-sola}

\begin{article}

\begin{opening}

\title{Observations of Coronal Mass Ejections with the Coronal Multichannel Polarimeter}

\author{H.~\surname{Tian}$^{1}$\sep
        S.~\surname{Tomczyk}$^{2}$\sep      
        S.\surname{W. McIntosh}$^{2}$\sep
        C.~\surname{Bethge}$^{2}$\sep
        G.~\surname{de Toma}$^{2}$\sep
        S.~\surname{Gibson}$^{2}$
       }

\runningauthor{H. Tian et al.}
\runningtitle{CoMP Observations of CMEs}

  \institute{ $^{1}$Harvard-Smithsonian Center for Astrophysics, Cambridge, MA, USA \\
                        e-mail:  \href{hui.tian@cfa.harvard.edu}{hui.tian@cfa.harvard.edu}\\
                 $^{2}$High Altitude Observatory, National Center for Atmospheric Research, Boulder, CO, USA \\
                        e-mail: \href{tomczyk@ucar.edu}{tomczyk@ucar.edu} 
             }

\begin{abstract}
The Coronal Multichannel Polarimeter (CoMP) measures not only the polarization of coronal emission, but also the full radiance profiles of coronal emission lines. For the first time, CoMP observations provide high-cadence image sequences of the coronal line intensity, Doppler shift and line width simultaneously over a large field of view. By studying the Doppler shift and line width we may explore more of the physical processes of the initiation and propagation of coronal mass ejections (CMEs). Here we identify a list of CMEs observed by CoMP and present the first results of these observations. Our preliminary analysis shows that CMEs are usually associated with greatly increased Doppler shift and enhanced line width. These new observations provide not only valuable information to constrain CME models and probe various processes during the initial propagation of CMEs in the low corona, but also offer a possible cost-effective and low-risk means of space weather monitoring.
\end{abstract}


\end{opening}

\section{Introduction}

Coronal mass ejections (CMEs) are probably the most important sources of adverse space weather effects (\textit{e.g.}, \opencite{Gosling1991}; \opencite{Gopalswamy2001}; \opencite{Wang2002}; \opencite{Howard2006}; \opencite{Zhang2007}; \opencite{Temmer2010}) and they are often associated with dramatic changes of coronal magnetic fields (\textit{e.g.}, \opencite{Zhang2005}; \opencite{Liu2009}; \opencite{Su2012}). Using mainly white-light coronagraphs, observations of CMEs are now made routinely both on the ground and in space. These instruments measure the polarized or total brightness of the corona and CMEs are usually identified as large-scale disturbances in the coronal intensity image sequences. White-light coronagraphs such as MK4 (\opencite{Elmore2003}; \opencite{Reiner2003}; \opencite{Gibson2006}) at the Mauna Loa Solar Observatory (MLSO) and the Large Angle Spectrometric Coronagraph (LASCO: \opencite{Brueckner1995}) onboad the Solar and Heliospheric Observatory (SOHO) have made great contributions to our understanding of the initiation and propagation of CMEs.

Spectroscopic observations of emission lines could provide valuable information on the plasma properties and dynamics in CMEs near the Sun (\textit{e.g.},\opencite{Harrison2003}; \opencite{Harra2003}; \opencite{Ko2003}; \opencite{Lin2005}; \opencite{McIntosh2010}; \opencite{Landi2010}; \opencite{Tian2012a}; \opencite{Giordano2013}). However, conventional slit spectrographs such as the Extreme-Ultraviolet Imaging Spectrometer (EIS: \opencite{Culhane2007}) onboard Hinode and the Ultraviolet Coronagraph Spectrometer (UVCS: \opencite{Kohl1995}) onboard SOHO can only observe a small region. In addition, repeated raster scans of the same region can only be done at a low cadence (\textit{e.g.}, about five minutes for Hinode/EIS, see \opencite{Tian2012a}) because it takes minutes or even hours to scan the region. A filter instrument, on the other hand, can provide high-cadence observations of a large field of view in the solar corona, thus offering significant advantages over a spectrograph when observing large-scale solar eruptions like CMEs. The LASCO-C1 instrument, which has revealed some important characteristics of CMEs' propagation in the low corona (\textit{e.g.}, \opencite{Plunkett1997}; \opencite{Schwenn1997}; \opencite{Zhang2001}), might be considered as one of such instruments. However, it took minutes to record a complete line profile \cite{Mierla2005} so that the cadence could not be high. Moreover, this instrument only lasted for less than two years during solar minimum, making it not very useful for CME studies.

The Coronal Multichannel Polarimeter (CoMP: \opencite{Tomczyk2008}) is also such an instrument. It uses a narrow-band tunable filter to take high-cadence observations of the polarization state at a few spectral locations across the profiles of three infrared lines (Fe~{\sc{xiii}} 1074.7 nm and 1079.8 nm, He~{\sc{i}} 1083.0 nm). Images taken by CoMP have a field of view (FOV) of 1.05 - 1.40 solar radii, a spatial resolution of 4.46$^{\prime\prime}$ pixel$^{-1}$, and a typical cadence of 30 s. The instrument was initially deployed at the Sacramento Peak of the National Solar Observatory in 2004. Several successful observations of coronal Alfv\'en waves \cite{Tomczyk2007,Tomczyk2009} and coronal cavities \cite{Schmit2009,Dove2011} have been performed since then. The instrument was recently moved to Mauna Loa Solar Observatory (MLSO) and started to obtain almost daily routine observations since October 2010.

Since CoMP can provide simultaneous measurements of the coronal line intensity, Doppler shift, line width, linear/circular polarization, and coronal density, it opens a completely new window for observations of the solar corona and CMEs. Here we report the first results of CoMP observations of CMEs. These observations might bring new insights into the initiation process of CMEs.

\section{Data Reduction and Correction}

Here we mainly use the three-point (sampled at three spectral locations 1074.50 nm, 1074.62 nm, 1074.74 nm) data of the Fe~{\sc{xiii}} 1074.7 nm line taken after December 2011. In these observations, sequences of the polarization (Stokes I, Q, and U only) images were obtained at each of these three spectral locations at a cadence of approximately 30 seconds. In this paper we mainly focus on the intensity (Stokes-I ) data.

Although there are measurements at only three spectral locations, the intensity profile is in principle the same as the spectral line profile obtained by spectrographs. Thus, we can simply apply a least-squares single Gaussian fit to each intensity profile and obtain the line-center intensity, central wavelength, and line width \cite{Tomczyk2007,Tomczyk2009}. However, it takes too much time to apply the Gaussian fit to all intensity profiles in the full FOV for hundreds of frames. Fortunately, we found that a simple analytical solution can be derived from the three-point measurement. It is far more efficient to derive the line parameters by using the analytical solution. 

If $I_{1}$, $I_{2}$, and $I_{3}$ are the measured intensities at known wavelengths (spectral locations) $\lambda_{1}$, $\lambda_{2}$, and $\lambda_{3}$, we have the following set of three equations:

\begin{equation}
\emph{$I_{1}=i\mathrm{e}^{\frac{-(\lambda_{1}-\lambda_{0})^2}{w^2}}$}, 
\end{equation}

\begin{equation}
\emph{$I_{2}=i\mathrm{e}^{\frac{-(\lambda_{2}-\lambda_{0})^2}{w^2}}$}, 
\end{equation}

\begin{equation}
\emph{$I_{3}=i\mathrm{e}^{\frac{-(\lambda_{3}-\lambda_{0})^2}{w^2}}$},
\end{equation}

It is clear that there are only three unknowns: the line center intensity [$i$], center wavelength [$\lambda_{0}$], and exponential line width [$w$]. Therefore, we should be able to directly derive these three unknowns from the three equations. If we take the natural logarithm of the ratios $I_{3}/I_{2}$ and $I_{1}/I_{2}$ and denote them as $a$ and $b$, we have:

\begin{equation}
\emph{$a=\mathrm{ln}(\frac{I_{3}}{I_{2}})=\frac{-(\lambda_{3}-\lambda_{0})^2}{w^2}+\frac{(\lambda_{2}-\lambda_{0})^2}{w^2}$}, 
\end{equation}

\begin{equation}
\emph{$b=\mathrm{ln}(\frac{I_{1}}{I_{2}})=\frac{-(\lambda_{1}-\lambda_{0})^2}{w^2}+\frac{(\lambda_{2}-\lambda_{0})^2}{w^2}$},
\end{equation}

If we take the sum $a+b$ and define $\delta$ as the spectral pixel size ($\lambda_{2}$ - $\lambda_{1}$ or $\lambda_{3}$ - $\lambda_{2}$), the line width can be derived and expressed as

\begin{equation}
\emph{$w=\sqrt{\frac{-2\delta}{a+b}}$}.
\end{equation}

The Doppler shift relative to $\lambda_{2}$ can be derived by taking the difference $a-b$:

\begin{equation}
\emph{$v=\lambda_{0}-\lambda_{2}=\frac{w^2}{4\delta}(a-b)$}.
\end{equation}

Finally, the line center intensity can be computed as

\begin{equation}
\emph{$i=I_{2}\mathrm{e}^{\frac{v^2}{w^2}}$}.
\end{equation}

 \begin{figure} 
 \centerline{\includegraphics[width=0.9\textwidth]{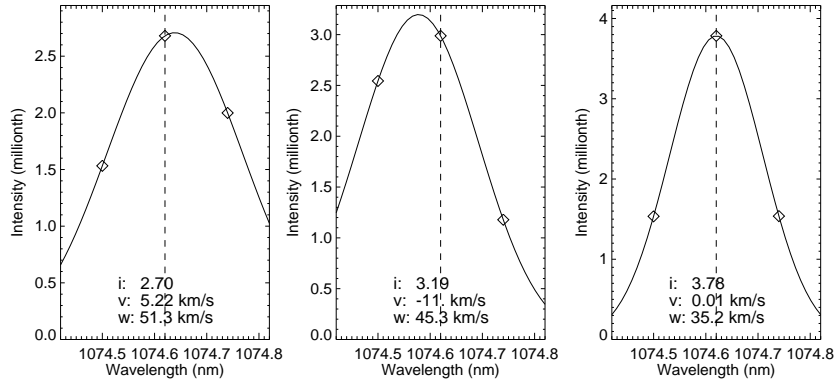}}
 \caption{Three examples of observed emission line profiles (diamonds) and the analytical solution (solid line). The dashed line indicates the rest wavelength of the line. The line center intensity [$i$], Doppler shift [$v$] and line width [$w$] are marked in each panel. }\label{f1}
\end{figure}

We show in Figure~\ref{f1} three examples of observed intensity profiles and the line parameters derived by using the analytical solution. The solid line in each panel is the Gaussian function constructed by using the derived line parameters. Here a positive and negative values of the Doppler shift mean red shift (away from the Earth) and blue shift (towards the Earth), respectively. We do not remove the instrument filter width from the measured line width since the filter width is a fixed value, and we are only interested in changes of the line width.

Once we derive the line parameters for each intensity profile in the FOV, we can produce maps of intensity, Doppler shift, and line width. Maps generated from observations at different times can then be used to make movies of intensity, Doppler shift, and line width. 
 
The Doppler-shift maps usually show predominant blue shift at the east limb and red shift at the west limb, which suggests an east-west trend in the line of sight (LOS) Doppler shift. This trend is at least partly caused by the rotation of the solar corona. We calculate the median value of Doppler shift at each Solar-X location to produce the east - west trend (Doppler shift as a function of Solar-x). A median filter is then applied to this trend to eliminate possible abnormal values. Then we apply a fifth-order polynomial fit to the filtered trend. The resulting smooth trend is then subtracted from the map of Doppler shift.

Since there is no calibration lamp or cold lines, we could not perform an absolute wavelength calibration. We simply assume that the median value of Doppler shift is zero in each image. This assumption is usually valid since non-radial flows should on average cancel each other out (\textit{e.g.}; \opencite{Hassler1999}; \opencite{Peter1999}; \opencite{Tian2010}) at the limb and CoMP has such a large FOV. In addition, we are only interested in large Doppler shift perturbation which is unlikely to be affected by the accuracy of the absolute wavelength scale.

\section{First Results of CME Observations}

As we mentioned previously, CoMP provides simultaneous high-cadence (30 seconds) observations of coronal line intensity, Doppler shift, and line width in a large FOV for the first time.  Such completely new types of  observations may provide new insights into the processes of CME initiation and propagation. We have checked the CoMP data archive as well as the Atmospheric Imaging Assembly (AIA: \opencite{Lemen2012}) onboard the Solar Dynamics Observatory (SDO) and SOHO/LASCO data, and found 27 clear cases where CMEs or CME-related signatures were observed by CoMP between December 2011 and February 2013. Table~\ref{t1} lists some information (observation date, approximate time when CoMP observes the CME, associated flare class, east/west limb, white light signatures and other characteristics) of these CMEs. The data of flare class is taken from Hinode Flare Catalogue (\href{http://st4a.stelab.nagoya-u.ac.jp/hinode\_flare/}{http://st4a.stelab.nagoya-u.ac.jp/hinode\_flare/}) and NOAA/SWPC Solar Region Summary (\href{http://www.swpc.noaa.gov/ftpmenu/warehouse.html}{http://www.swpc.noaa.gov/ftpmenu/warehouse.html}).

 \begin{table*}[]
\caption[]{CMEs observed by CoMP between December 2011 and February 2013. }\label{t1}
\begin{center}
\begin{tabular}{p{0.32cm}  p{1.7cm}  p{0.7cm}  p{0.8cm}  p{0.66cm} p{2.1cm} p{2.9cm}}
\hline ID & Date & Time & Flare Class &  Limb & White Light Signature & Remarks \\
\hline 
 1 & 07 Dec 2011 & 20:10 & C2.7 & east & no & loop oscillation  \\
 2 & 30 Dec 2011 & 20:18 & C4.4 & west & MK4 \& LASCO & flux rope, Type-II burst  \\
 3 & 09 Jan 2012 & 20:13 &  C2.6 & east & LASCO & \\
 4 & 14 Jan 2012 & 21:12 & C2.1?  & west & no &  \\
 5 & 15 Mar 2012& 01:33 & C1.1  & west & MK4 \& LASCO &  prominence eruption \\
 6 & 17 Mar 2012& 23:55 & B8.1 & west & MK4 \& LASCO & \\
 7 & 27 Mar 2012& 21:40 & no & west & MK4 \& LASCO & \\
 8 & 11 Apr 2012 & 20:40 & no & west & MK4 \& LASCO & \\
 9 & 27 Apr 2012 & 17:20 & no & east & MK4 \& LASCO & \\
 10 & 15 May 2012 & 17:20 & no & west & MK4 \& LASCO & null reconnection \\
 11 & 26 May 2012 & 20:38 & no &  west & partial halo in MK4 \& LASCO & originates from back side of the Sun \\
 12 & 01 Jun 2012 & 22:16 & C3.3 & west &  MK4 \& LASCO & Type-II burst \\
 13 & 08 Jun 2012 & 23:24 & no & west & MK4 \& LASCO & prominence eruption \\
 14 & 06 Jul 2012 & 23:05 &  X1.1 & west &  MK4 \& LASCO & Type-II burst \\
 15 & 08 Jul 2012 & 19:00 & no & west & MK4 \& LASCO & slowly rising loops \\
 16 & 13 Jul 2012 & 19:45 &  C1.3 & west &  no & \\
 17 & 07 Aug 2012 & 19:02 &  C3.0 & east & no & prominence eruption \\
 18 & 16 Aug 2012 & 00:30 & no & east & MK4 \& LASCO & prominence eruption \\
 19 & 24 Aug 2012 & 19:30 & no & east & LASCO & narrow eruption \\
 20 & 15 Sept 2012 & 22:57 & B9.6 & west & faint in LASCO & Type-II burst \\
 21 & 22 Sept 2012 & 19:30 & B9.2 & east & MK4 \& LASCO & \\
 22 & 01 Nov 2012 & 21:56 & no & west & no &\\
 23 & 12 Nov 2012 & 18:55 & C2.0 & east & LASCO &\\
 24 & 08 Jan 2013 & 19:11 & C4.0 & west & no & prominence eruption \\
 25 & 01 Feb 2013 & 19:23 & no & west & LASCO &\\ 
 26 & 04 Feb 2013 & 20:43 & C1.9 & east & LASCO &\\
 27 & 15 Feb 2013 & 20:00 & no & east & LASCO &\\
\hline
\end{tabular}
\end{center}
\end{table*}

Inspection of the movies shows that the typical characteristics of CMEs in the CoMP data are the dramatic changes of Doppler shift and the obviously enhanced line width. Although all CMEs do show some perturbation in the image sequences of the intensity, the perturbation is much more obvious in the image sequences of  the Doppler shift and line width. This is not difficult to understand since higher-order moments are usually more sensitive to changes. The significant changes of Doppler shift are probably largely associated with the coronal response to the mass eruptions and lateral expansion of CMEs. The increased line width might be caused by the enhanced flow inhomogenuity and turbulence in various substructures of CMEs in the LOS direction.

We have to mention that the values of Doppler shift during CME eruption should not be interpreted as the line-of-sight component of the CME propagation speed in these observations. The reason is that the line wings are poorly sampled in our three-point measurements. However, the red and blue patterns in the Dopplergram do suggest that the plasma motions averaged along the line of sight direction are directed away and towards the observer, respectively. These motions include not only the outward movement of the CME, but also its lateral expansion and the dynamic response of the surrounding coronal plasma along the line-of-sight direction. This explains the complex Doppler-shift patterns in some CME eruption events. Nevertheless, the Doppler shift should experience a dramatic change in all CME eruption events and the largely perturbed (usually enhanced) Doppler shift should have an outward motion as the CME moves outward. Thus, the outward motion of largely enhanced Doppler shift is a good marker of CME eruption. In the future we plan to perform measurements of the line profile at more spectral locations. A better spectral sampling may help to separate these different types of motions due to their different speeds.

 \begin{figure} 
 \centerline{\includegraphics[width=0.9\textwidth]{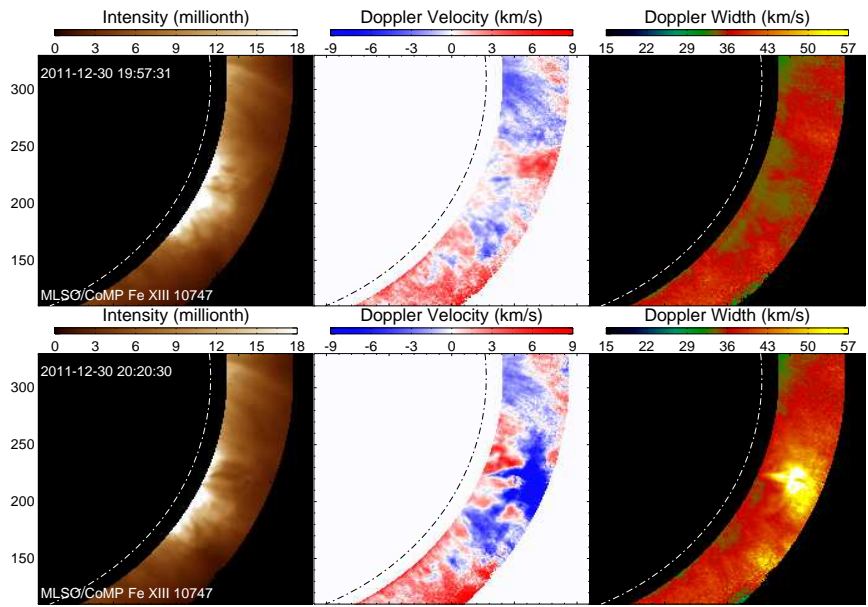}}
 \caption{An eruption observed by CoMP on 30 December 2011. The first and second rows show images of the line center intensity, Doppler shift, and line width at 19:57:31 (before eruption) and 20:20:30 UT (during eruption), respectively. The dot - dashed line in each panel marks the limb of the solar disk. Two ESM  movies (m2a.mov, m2b.mov) are associated with this figure. }\label{f2}
\end{figure}

Figure~\ref{f2} shows the three line parameters before and during the 30 December 2011 CME.  The CME-caused changes of all line parameters can be clearly identified through a comparison between the first and second rows. Continuous evolution of different line parameters can be seen from the Electronic Supplementary Material (ESM) movies m2a.mov (full FOV) and m2b.mov (partial FOV). The outward propagating ejecta causes a dimming of $\sim$ 50\% in the intensity, shifts the line center by about 20 km s$^{-1}$ blueward, and enhances the line width by $\sim$ 20 km s$^{-1}$. The spatial pattern of the intensity change roughly coincides with those of the Doppler shift and line width enhancement. 

From LASCO movies (\href{http://lasco-www.nrl.navy.mil/daily\_mpg/}{http://lasco-www.nrl.navy.mil/daily\_mpg/}), we can see that this CME developed into a ring-shaped propagating feature in the FOV of LASCO-C2 and C3. Such a feature is likely to be a signature of flux rope or magnetic cloud (\textit{e.g.}, \opencite{DeForest2011}). From Figure~\ref{f2} we can see that this CME erupts from an active region (AR). It is not clear whether the flux rope existed prior to eruption (\textit{e.g.}, \opencite{Gibson2006}) or formed during the eruption (\textit{e.g.}, \opencite{Cheng2011}). 

A possible flux rope is also identified from the LASCO-C2 and C3 movies on 13 October 2011. From AIA images, we can see that this flux-rope-type CME seems to originate from a coronal cavity at the northwest limb. Such a connection favors the flux-rope interpretation of coronal cavities \cite{Low1995,Gibson2006}. Unfortunately, on that day CoMP data was only available prior to the CME eruption and the observation of the cavity is limited by an obstruction in the instrument. We also checked the list of cavities of \opencite{Forland2011} but none of the erupting cavities were caught by CoMP. We hope that future observations may catch the complete process of flux rope ejection and thus better our understanding of the role played by flux ropes in CME initiation and eruption.

 \begin{figure} 
 \centerline{\includegraphics[width=0.9\textwidth]{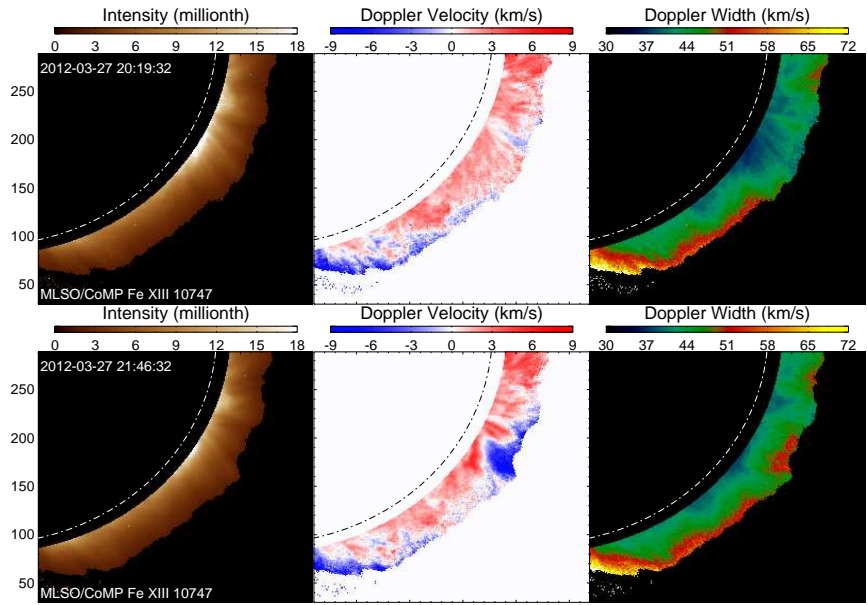}}
 \caption{An eruption observed by CoMP on 27 March 2012. The first and second rows show images of the line-center intensity, Doppler shift, and line width at 20:19:32 (before eruption) and 21:46:32 UT (during eruption), respectively. The dot-dash line in each panel marks the limb of the solar disk. An ESM movies (m3.mov) is associated with this figure. }\label{f3}
\end{figure}

The intensity perturbation of the 27 March 2012 CME is very small, which might be due to the fact that the CME propagates largely off the plane of sky (POS). Despite the weak signal in the intensity, we see an obvious change of the Doppler shift and line width. From Figure~\ref{f3} and the ESM movie m3.mov we can clearly see the significant perturbation of Doppler shift and line width as the CME propagates into the FOV of CoMP.

 \begin{figure} 
 \centerline{\includegraphics[width=0.9\textwidth]{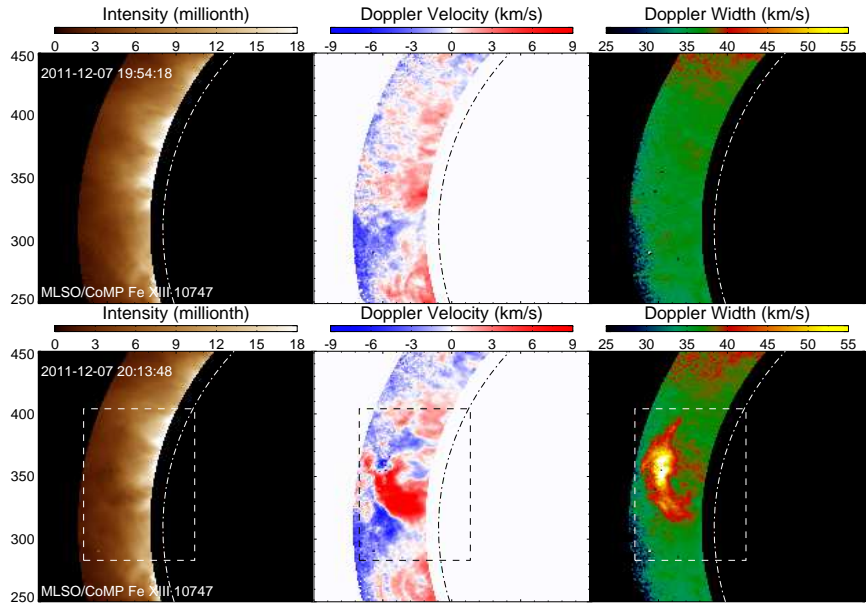}}
 \caption{An eruption observed by CoMP on 7 December 2011. The first and second rows show images of the line-center intensity, Doppler shift, and line width at 19:54:18 and 20:13:48 UT, respectively. The dot - dashed line in each panel marks the limb of the solar disk. The rectangular region indicates the field of view shown in Figure~\ref{f5}. An ESM movie (m4.mov) is associated with this figure. }\label{f4}
\end{figure}

 \begin{figure} 
 \centerline{\includegraphics[width=0.9\textwidth]{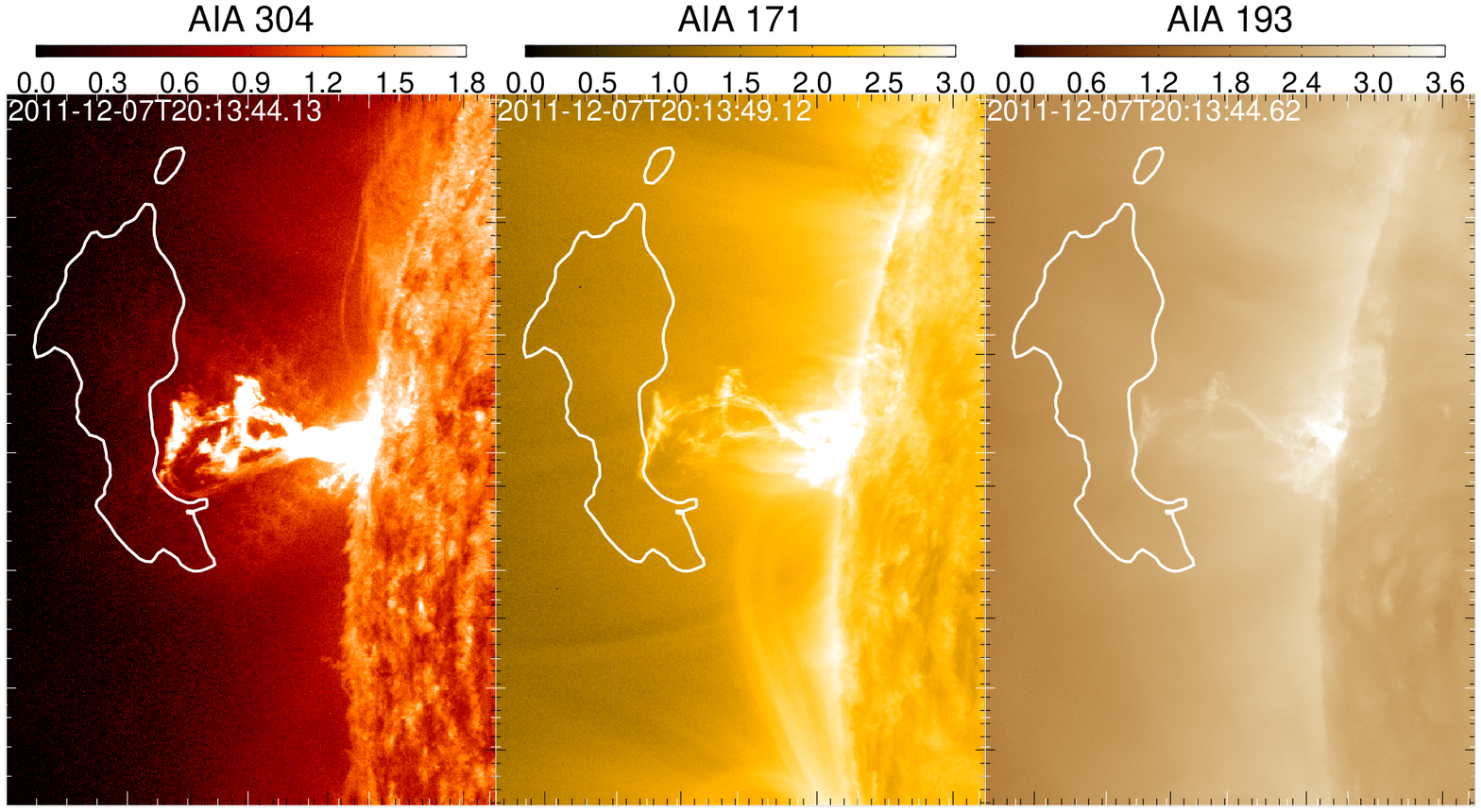}}
 \caption{AIA observations of the 7 December 2011 event. Images of the AIA 304~\AA, 171~\AA~and 193~ \AA~passbands around 20:13:48 are presented from left to right. The contours mark locations where the line width of Fe~{\sc{xiii}} 1074.7 nm is larger than 42 km s$^{-1}$. }\label{f5}
\end{figure}

The 7 December 2011 eruption only revealed very weak perturbation in the daily LASCO-C2 and C3 movies. However, as we can see from Figure~\ref{f4} and Figure~\ref{f5}, this eruption was clearly recorded by both CoMP and SDO/AIA. The dominant ions in the 304~\AA, 171~\AA and 193~\AA ~passbands are He~{\sc{ii}}, Fe~{\sc{ix}}, and Fe~{\sc{xii}}, respectively \cite{ODwyer2010}. The eruption appears as a dark propagating feature in the CoMP intensity data, as can be seen from Figure~\ref{f4} and the ESM movie m4.mov. The eruption also shifts the line center by about 20 km s$^{-1}$ redward and enhances the line width by more than 20 km s$^{-1}$. The dark feature roughly coincides with the enhanced line width and largely perturbed Doppler shift. 

In Figure~\ref{f5} we present the high-resolution AIA images in three passbands. These images were taken around 20:13:48 UT, the time when the images in the lower panel of Figure~\ref{f5} were taken. The most enhanced line width (larger than 42 km s$^{-1}$) region, which is also approximately the region where the most significant perturbation of the line center intensity and Doppler shift  occurs, is outlines by the white contours in each panel. It seems that this region is immediately ahead of the ejecta. A comparison between Figure~\ref{f4} and Figure~\ref{f5} suggests that the ejecta are also associated with large changes of Doppler shift and enhancement of the line width. However, these changes are clearly not as prominent as those ahead of the ejecta. The large perturbation ahead of the ejecta is likely largely caused by the blastic disturbance of the plasma and magnetic environment around the leading edge of the fast ejection. 

Interestingly, we do not see any obvious dark propagating feature ahead of this ejecta in the AIA movies (not shown here).  It is known that all of these three passbands of AIA have significant contribution from emission of cooler materials, whereas the CoMP emission is almost purely from the hot Fe~{\sc{xiii}} ion. But it is still not clear how the perturbation ahead of the ejecta causes a reduction in the intensity of Fe~{\sc{xiii}} 1074.7 nm but no reduction in the AIA intensities.  

Another interesting aspect of the 7 December 2011 eruption is the transverse oscillation of a loop system excited by the ejecta. The loop oscillation is best seen in the AIA 171\AA ~movie (not shown here). The oscillating loop is clearly present in Figure~\ref{f5} and is about 432 Mm in length. The oscillation has a period of about 15 minutes and lasts for four cycles before damping out. The kink speed (twice the loop length divided by the period) at the apex of this oscillating loop can thus be calculated as 960 km s$^{-1}$. The Alfv\'en speed inside the oscillating loop can be constrained to the range of 960/$\sqrt{2}$ km s$^{-1}$-960 km s$^{-1}$ (\textit{e.g.}, \opencite{Edwin1983}; \opencite{Aschwanden1999}; \opencite{Nakariakov2001}; \opencite{Wang2004}; \opencite{VanDoorsselaere2008}; \opencite{Chen2011}; \opencite{Feng2011}; \opencite{Tian2012b}; \opencite{White2012}). The two Fe~{\sc{xiii}} lines used by CoMP (1074.7 nm \& 1079.8 nm) can in principle be used to diagnose the coronal electron density. Combining information of the Alfv\'en speed and electron density, we can directly calculate the coronal magnetic field strength. Unfortunately, this oscillating loop is barely resolved by CoMP. In addition, the density diagnostics of CoMP are still not finalized. We are planning to do joint observations of Hinode/EIS and CoMP, trying to use the density diagnostics of EIS to calibrate those of CoMP.

 \begin{figure} 
 \centerline{\includegraphics[width=0.9\textwidth]{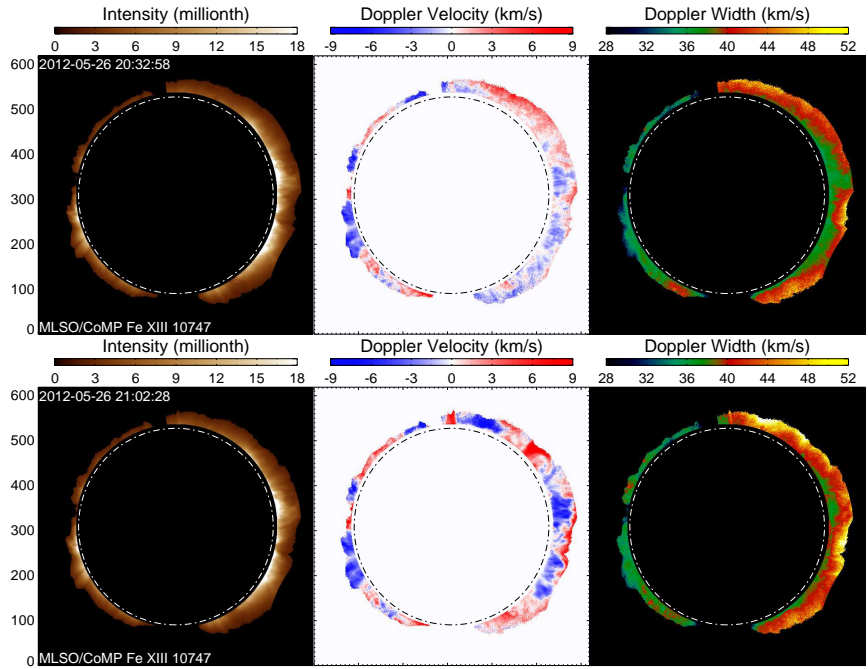}}
 \caption{A partial halo-CME observed by CoMP on 26 May 2012. The first and second rows show images of the line center intensity, Doppler shift and line width at 20:32:58 and 21:02:28 UT, respectively. The dot - dashed line in each panel marks the limb of the solar disk.}\label{f6}
\end{figure}

Figure~\ref{f6} shows snapshots of CoMP observations of a partial-halo CME. A complete halo is seen at the viewpoint of STEREO-A. This eruption seems to originate from the backside of the Sun and propagates away from the Sun. The quickly developed large-scale disturbance can be clearly identified from movies of the Doppler shift and line width (available on \href{http://mlso.hao.ucar.edu/mlso\_datasum\_comp.php?2012\&5\&26\&COMP}{http://mlso.hao.ucar.edu/mlso\_datasum\_comp.php?2012\&5\&26\&COMP}, mainly at the west limb). Figure~\ref{f6} clearly shows that the perturbation in the intensity data is not as significant as in the Doppler shift and line width, probably because of the strong foreground and background coronal emission. The quickly-developed perturbation might be associated with the propagation of \texttt{"}EUV waves\texttt{"}. \opencite{Chen2010} suggests that spectroscopic observations may be used to identify the nature of \texttt{"}EUV waves\texttt{"}, namely fast waves (\textit{e.g.}, \opencite{Patsourakos2009},\opencite{Shen2012}) or non-wave phenomenon such as successive stretching of magnetic field lines (\textit{e.g.}, \opencite{Chen2002}). Thus, combining numerical simulations and the spectroscopic-like large-FOV CoMP observations might reveal more insights into the process of \texttt{"}EUV waves\texttt{"}.

CoMP observations of the 26 May 2012 CME suggest the importance of monitoring space weather at ground. The 26 May 2012 CME first appears as a partial halo in the FOV of LASCO-C2 at 20:57, which is about 20 minutes later than the time (20:38 UT) when CoMP observes the larger-scale perturbation in the coronal line width and Doppler shift. So it is clear that CoMP observations can be very important for the prediction of Earth-directed halo CMEs. In this sense, observations of halo-CMEs by ground-based instruments like CoMP provide a cost-effective and low-risk means of space weather monitoring.

\section{Summary and Future Perspectives}

CoMP provides high-cadence large-FOV spectroscopic-like observations of CMEs for the first time. We have presented first results of CME observations by CoMP in this paper. Our results show that the primary characteristics of CMEs are the dramatic change of the Doppler shift and obviously enhanced line width. The information provided by these CoMP observations might be important for our better understanding of various CME processes such as the formation of flux ropes and propagation of \texttt{"}EUV waves\texttt{"}. CoMP observations clearly demonstrate that space-weather monitoring is possible with inexpensive and low-risk observations from the ground. So far we have identified 27 obvious CMEs in the CoMP data. With the increase of the solar activity, we expect to observe more CMEs in the future. 

CoMP observations of CMEs are not restricted to the images of line center intensity, Doppler shift, and line width. In fact CoMP measures the complete polarization state of three emission lines. The linear (Stokes-Q and U) and circular (Stokes-V) polarization data are not used in our analysis because our preliminarily processed polarization data shows no significant change of the linear polarization during CME eruptions and the circular-polarization signal is weak.  We plan in the future to examine these data more thoroughly in order to establish whether changes may be evident if the data are averaged over longer time intervals, or whether a more optimal observation program might be established for analyzing CMEs with linear/circular polarization.  Linear polarization is a promising diagnostic of coronal magnetic structures (\textit{e.g.}, \opencite{Rachmeler2013}) and it may be possible to demonstrate changes in magnetic topology by examining data before, during and after CME eruptions.

As we mentioned above, density diagnostics using the two Fe~{\sc{xiii}} lines is underway. Once this is finalized, we should be able to study density changes during CMEs, which could be used to estimate the CME mass (\opencite{Harrison2003}; \opencite{Jin2009}; \opencite{Tian2012a}). In addition, combining the density diagnostics and the Alfv\'en wave (Doppler shift oscillation) analysis \cite{Tomczyk2007,Tomczyk2009,McIntosh2008}, we can produce maps of the electron density, Alfv\'en speed, and magnetic-field strength before and after CMEs. Such information can be used to constrain background coronal parameters prescribed in models of CMEs. 

CoMP is just a prototype of the Large-aperture Coronagraph component of the proposed COronal Solar Magnetism Observatory (COSMO). With a FOV of 1.05 - 2.0 solar radii, COSMO can greatly expand the ability of CME observations by CoMP. In addition, the Chromosphere and Prominence Magnetometer (ChroMag) component of COSMO can monitor solar activity on the disk. Thus, a combination of both components is likely to monitor the complete process of (halo) CME initiation and early-phase propagation. 

%

%
\begin{acks}
SDO is the first mission of NASAÕs Living With a Star (LWS) Program. H. Tian's work at CfA is supported  under contract 8100002705 from Lockheed-Martin to SAO. Part of this work was done at NCAR, where H. Tian was supported under the ASP Postdoctoral Fellowship Program. The National Center for Atmospheric Research is sponsored by the National Science Foundation. We thank L. Sitongia for processing the CoMP level-0 data. We also thank P. Judge, B.C. Low, and H. Peter for helpful discussions. 
\end{acks}

%
%
\bibliography{ms.bib}

\end{article} 
\end{document}